\documentclass[10pt,journal]{IEEEtran}
\IEEEoverridecommandlockouts

\newtheorem{problem}{Problem}
\newtheorem{theorem}{Theorem}
\newtheorem{corollary}{Corollary}
\newtheorem{lemma}{Lemma}
\newtheorem{remark}{Remark}
\newtheorem{definition}{Definition}
\newtheorem{proposition}{Proposition}
\newtheorem{example}{Example}
\newtheorem{assumption}{Assumption}

\usepackage{amsmath,amsfonts,amssymb,color}
\usepackage{epsfig}
\usepackage{psfrag}
\usepackage{algorithm}
\usepackage{algpseudocode}
\usepackage{array}
\usepackage{epstopdf}
\usepackage{tikz, cite}
\usepackage{multicol,multirow}
\usepackage{algorithmicx}
\usepackage{float}
 
\DeclareMathOperator*{\argmax}{arg\,max}

\title{\LARGE \bf Algorithms for Influence Maximization in Socio-Physical Networks}
\author{Hemant Gehlot, Shreyas Sundaram, and Satish V. Ukkusuri % <-this % stops a space
\thanks{Hemant Gehlot and Satish V. Ukkusuri are with the Lyles School of Civil Engineering at Purdue University. Email: {\tt \{hgehlot,sukkusur\}@purdue.edu}. Shreyas Sundaram is with the School of Electrical and Computer Engineering at Purdue University. Email: {\tt sundara2@purdue.edu}. }%. 
}

\begin{document}
\maketitle
\thispagestyle{empty}
\pagestyle{empty}

% \documentclass{ifacconf}

% \usepackage{graphicx}      % include this line if your document contains figures
% \usepackage{natbib}        % required for bibliography
% \usepackage{amsmath,amsfonts,amssymb,color}
% \usepackage{epsfig}
% \usepackage{psfrag}
% \usepackage{algorithm}
% \usepackage{algpseudocode}
% \usepackage{array}
% \usepackage{epstopdf}
% \usepackage{tikz}
% \usepackage{natbib}
% \usepackage{float}
% \DeclareMathOperator*{\argmax}{arg\,max}

% \newenvironment{proof}{\textbf{Proof.}}{\hfill $\square$\\\\}
% \newcommand\numberthis{\addtocounter{equation}{1}\tag{\theequation}}
% \begin{document}
% \begin{frontmatter}
\begin{abstract}
Given a directed graph (referred to as \textit{social network}), the \textit{influence maximization} problem is to find $k$ nodes which, when influenced (or \textit{activated}), would maximize the number of remaining nodes that get activated under a given set of activation dynamics. 
In this paper, we consider a more general version of the problem that includes an additional set (or layer) of nodes that are termed as \textit{physical nodes}, such that a node in the social network is \textit{covered} by one or more physical nodes. A physical node exists in  one of two states at any time,  \textit{opened} or \textit{closed}, and there is a constraint on the maximum number of physical nodes that can be opened. In this setting, an inactive node in the social network becomes active if it has a sufficient number of active neighbors in the social network \textit{and} if it is covered by at least one of the opened physical nodes. This problem arises in scenarios such as disaster recovery, where a displaced social group (an inactive social node) decides to return back after a disaster (switches to active state) only after a sufficiently large number of groups in its social network return back and some infrastructure components (physical nodes) in its neighborhood are repaired (brought to the open state). 
We first show that this general problem is NP-hard to approximate within any constant factor. We then consider instances of the problem when the covering function between the social nodes and the physical nodes is bijective and characterize optimal and approximation algorithms for those instances.    
\end{abstract}

\section{Introduction}
The influence of social networks on the decision making of individuals has been documented in various studies in social science and economics \cite{brown1987social,kempe2003maximizing}. For example, it has been observed that people are more likely to adopt new products if their friends or relatives adopt them \cite{brown1987social}. It has also been found that the cumulative effect of people returning back in the neighborhood of a household after a disaster has a significant impact on a household's decision to return back \cite{li2010katrina,patterson2010role}. 
The paper \cite{kempe2003maximizing} formulated \textit{influence maximization} as a discrete optimization problem of finding the $k$ most influential nodes (also referred to as \textit{seed} nodes) in a social network, which on being influenced (or \textit{activated}) would maximize the spread of influence across the network. Since then, there have been several works that have focused on the influence maximization problem and its extensions \cite{chen2010scalable,goyal2011celf++,mossel2010submodularity,zhao2015relative,zhu2018social,saritacc2016online,lu2012complexity,lu2011approximation,wang2016bharathi}. A common model for influence propagation is the \textit{linear threshold model}, where each node in the social network has a threshold and an inactive node becomes active if the sum of influence from all of its active neighbors is at least equal to its threshold \cite{kempe2003maximizing,chen2010scalable}. 
The paper \cite{kempe2003maximizing} considered randomized thresholds in the linear threshold model; for such cases, it has been shown that exactly computing the spread of influence for a given set of seed nodes under the linear threshold model is \#P-hard \cite{chen2010scalable}, implying that there is unlikely to exist an efficient algorithm to optimally solve that problem. Noting that it is possible to deterministically estimate the thresholds in the real-world through surveys \cite{lu2012complexity}, the papers \cite{lu2012complexity} and \cite{lu2011approximation} define the \textit{deterministic linear threshold model (DLTM)} where the thresholds of nodes are deterministic. It has been proved that the spread of influence for a given set of seed nodes can be computed in polynomial-time under \textit{DLTM} \cite{lu2012complexity} and therefore our focus in this paper will also be related to \textit{DLTM}.

In certain scenarios, such as after disasters, a displaced social group such as a household or a community may not decide to return to its home unless a sufficiently large number of groups in its social network return back {\it and} certain infrastructure components in its residential neighborhood are repaired \cite{green2007impediments}. Our goal in this paper is to capture such scenarios. Specifically, we extend the traditional influence maximization problem by considering an additional set (or layer) of nodes (apart from the nodes in the social network) that are termed as \textit{physical nodes}, where each physical node \textit{covers} one or more social nodes. We term the combined network consisting of social and physical nodes as a \textit{socio-physical} network. A physical node exists in one of two states at a time, \textit{opened} or \textit{closed}, where an opened (resp. closed) physical node represents a repaired (resp. damaged) infrastructure component in the context of disaster recovery. Therefore, a necessary condition for an inactive social node to become active is that it should be covered by at least one of the opened physical nodes. However, it might not be possible to repair all of the physical components with a given repair budget (e.g., repair crew and resources), as some of the components may fail permanently if they are not repaired in a timely manner \cite{gehlot2019optimal,gaspard2007impact}. Thus, we consider a constraint on the total number of physical nodes that can be opened (in addition to the constraint on the total number of seed nodes in the social network as in the influence maximization problem).  

For the standard influence maximization problem under \textit{DLTM}, the paper \cite{lu2011approximation} proved that the optimal solution is NP-hard to approximate within any constant factor for the case when some inactive nodes require more than one neighboring nodes to become active; however, there exists a constant factor approximation algorithm for instances of the problem where each inactive node can be activated by only one active neighboring node \cite{lu2011approximation}. Therefore, we focus on these tractable cases where an inactive node in the social network is activated when at least one of its neighboring nodes is activated, and is connected to at least one opened physical node. Since a social group such as a community or a household consists of multiple individuals, each social node has an associated weight and under the problem setting described above, we characterize optimal and near-optimal algorithms for special instances of the problem that maximize the total weight of the social nodes that get activated. There are high level similarities between our problem and cascades in multilayer networks \cite{salehi2015spreading,yaugan2012analysis}. The paper \cite{salehi2015spreading} provides a review of spreading processes in multilayer networks and \cite{yaugan2012analysis} analyzes a linear threshold model for multiplex networks. However, the existing studies do not focus on influence maximization under \textit{DLTM} where the spread of influence in the social network is also a function of the recovery of nodes in another network (i.e., the recovery of physical nodes in our problem).

\subsection*{Our contributions}
As mentioned before, we focus on the case where a necessary condition for an inactive node in the social network to become active is that at least one of its neighboring nodes should be active. Along with the aforementioned assumption, we focus on the case when there is a one-to-one mapping between the social and physical nodes. Under these assumptions, we first characterize a $\max\{\frac{e}{e-1},\frac{w^{max}}{w^{min}}\}$ approximation algorithm\footnote{For $\rho\ge 1$, a polynomial-time algorithm is said to be a $\rho$ approximation algorithm for a maximization problem if the value computed by the algorithm is at least $\frac{1}{\rho}$ times the optimal value.} when the social network is a general directed graph, where $w^{max}$ and $w^{min}$ are the largest and smallest weights of social nodes, respectively, and $e$ is the base of the natural logarithm. We then argue that the same algorithm has a better approximation ratio when  the social network is represented by a bipartite graph. 
Finally, under the two aforementioned assumptions, we characterize a polynomial-time optimal algorithm when the social network is a set of disjoint trees such that all the edges in a tree are directed away from the root node.      

This paper is organized as follows. Section II focuses on the problem statement. After this, we argue that the general problem is NP-hard to approximate within any constant factor. In the subsequent sections, we consider special cases of the general problem, and characterize optimal and approximation algorithms for them. 
Finally, we conclude the study and provide some future research directions.

\section{Problem statement}
We consider a scenario with two types of nodes, \textit{social nodes} and \textit{physical nodes}. In the context of disaster recovery, a social node could represent a social group such as a community or a household, and a physical node could represent an infrastructure component such as roads in an area or a portion of the power network. The set of social nodes is represented by $\mathcal{V}$, where $|\mathcal{V}|=N$. The weight of a social node $j\in\mathcal{V}$ is denoted by $w_j \in \mathbb R_{> 0}$; for example, this weight could represent the number of individuals in the corresponding social group. Every social node exists in one of two possible states at each point in time, \textit{active} or \textit{inactive}. Since individuals belonging to the same social group (such as a household or tight-knit community) are interconnected, and decisions (such as whether to return after a disaster) are often made collectively by the individuals in the group, we will use a single state to represent the states of all the individuals within a group \cite{patterson2010role,li2010katrina}.  
% In the disaster recovery scenario, an inactive social node could represent a household that is displaced after a disaster and has not yet decided to return back, whereas an active node could represent a household that has decided to return back to its home. 
The set of physical nodes is represented by $\mathcal{W}$, where $|\mathcal{W}|=M$. Every physical node exists in one of two possible states at each point in time, \textit{opened} or \textit{closed}.  

The relationships between the different social nodes are represented by a directed graph $G=\{\mathcal{V},\mathcal{E}\}$. An edge $(i,j)\in \mathcal{E}$ represents a directed edge starting from social node $i$ and ending in social node $j$; social node $i$ is an \textit{incoming} neighbor of social node $j$. There exists a mapping between the social and physical nodes such that each physical node \textit{covers} one or more social nodes. For all $l\in \mathcal{W}$, the (non-empty) set of social nodes that are covered by physical node $l$ is denoted by $\mathcal{Q}_l\subseteq \mathcal{V}$. Also, each social node is covered by at least one of the physical nodes, i.e., $\cup_{l\in \mathcal{W}} \mathcal{Q}_l = \mathcal{V}$. We assume that there are no edges present between the physical nodes.\footnote{We keep the analysis involving dependencies between physical nodes as a future avenue for research.}

We assume that time progresses in discrete time-steps capturing the resolution at which decisions are made by the social nodes to become active or not \cite{kempe2003maximizing}. We index time-steps by $t\in\mathbb{N}=\{0,1,2,\ldots\}$. The total number of social nodes that can be activated at time-step 0 due to the constraints on the budget is given by $K_s (\le N)$. The social nodes that are activated at time-step 0 are referred as the \textit{seed} nodes. In the context of disaster recovery, seed nodes could represent the households that are provided various incentives and aids such as transition assistance, accelerated tax returns, disaster housing assistance, etc., by government agencies  that help them to return back \cite{WhatItTa76:online,TaxCredi54:online}. We consider the \textit{progressive} case where an active social node does not switch back to the inactive state \cite{kempe2003maximizing}. There is a constraint $K_p (\le M)$ on the total number of physical nodes that are opened. The decision to open a physical node or not is taken at time-step 0. For physical nodes also, we consider the progressive case, i.e., once a physical node opens at time-step 0, it remains opened for all subsequent time-steps. In the context of disaster recovery, that assumption is justified because infrastructure components face accelerated deterioration after disasters \cite{gaspard2007impact}, so it can be assumed that normal deterioration processes do not significantly change the health of a component once it is repaired  \cite{gehlot2019optimal}. 

The interactions between the different social nodes are derived from the \textit{DLTM} \cite{lu2012complexity}. Specifically, denote the number of incoming neighbors (in graph $G$) of social node $j$ by $\eta_j$. Each social node $j\in \mathcal{V}$ has a threshold $\theta_j \in \mathbb Z_{> 0}$ such that $1\le \theta_j\le \eta_j$. Let $\eta_{j,t}$ be the number of active incoming neighbors (in graph $G$) of social node $j$ at time-step $t$. 
An inactive social node $j$ at time-step $t$ becomes active at time-step $t+1$ if the number of active incoming neighbors of node $j$ at time-step $t$ is at least equal to $\theta_j$ (i.e., $\eta_{j,t}  \ge \theta_j$), and in our setting, at least one of the physical nodes that covers social node $j$ is in the open state at time-step $t$. 
Note that the states of social nodes are guaranteed to reach a steady state after at most $N$ time-steps, because at least one social node gets activated in each time-step until the states stop changing. Therefore, we refer to the total weight of the social nodes that are activated by the end of time-step $N$ as the total weight of the social nodes that \textit{eventually} get activated. Under the assumptions and dynamics described above, we focus on the following problem.
\begin{problem}\label{problem} %HG: more description on edges between social and physical nodes like N\ge M and full covering of social groups by physical nodes?
Given a social network $G=\{\mathcal{V},\mathcal{E}\}$ of $N(\ge 1)$ social nodes with node weights $\{w_j\}$ and node thresholds $\{\theta_j\}$, and a set $\mathcal{W}$ of $M(\ge 1)$ physical nodes where the covering of social nodes by physical nodes is given by $\{\mathcal{Q}_l\}$, determine $K_s (\le N)$ social nodes that should be selected as the seed nodes and $K_p (\le M)$ physical nodes that should be opened in order to maximize the total weight of the social nodes that eventually get activated. 
\end{problem}

We first argue that Problem \ref{problem} is NP-hard to approximate within any constant factor (in general). After that, we will look at various special cases of this problem and characterize optimal/approximation algorithms to solve them.

\section{Inapproximability}
We first define an approximation algorithm \cite{cormen2009introduction}.
\begin{definition}[Approximation algorithm]
Let $C$ be the optimal value of a maximization problem and $C'$ be the value computed by a polynomial-time algorithm $A$. Then, $A$ is a $\rho$ approximation algorithm if $\frac{C}{C'}\le \rho$ for all the instances of the maximization problem.
\end{definition}

We now present the following inapproximability result.
\begin{proposition}
Problem \ref{problem} is NP-hard to approximate within a factor $N^{1-\epsilon}$ for any $\epsilon \in (0,1)$.\footnote{This means that there cannot be a $\rho$ approximation algorithm for Problem \ref{problem} such that $\rho \le N^{1-\epsilon}$ for any $\epsilon \in (0,1)$, unless P = NP.}
\end{proposition}

\begin{IEEEproof}
Consider the instances of Problem \ref{problem} where $K_p=M$ (i.e., all the physical nodes can be opened), $w_j=w, \forall j\in \mathcal{V}$ (i.e., the weights of all the social nodes are the same) and for all $j\in \mathcal{V}, \theta_j \le 2$ (i.e., each inactive social node requires one or two active incoming neighboring nodes to get activated). Since all the physical nodes can be opened, such instances of Problem \ref{problem} are equivalent to the instances of the influence maximization problem under \textit{DLTM}, which are NP-hard to approximate within a factor of $N^{1-\epsilon}$ for any $\epsilon\in (0,1)$, where $N$ is the number of social nodes (see Theorem 5 of \cite{lu2011approximation}). Thus, the result follows.
\end{IEEEproof}

Although the influence maximization problem under \textit{DLTM} is NP-hard to approximate within any constant factor when each inactive social node requires one or two active incoming neighboring nodes to get activated (i.e., for all $j\in \mathcal{V}, \theta_j \le 2$), the paper \cite{lu2011approximation} showed that the problem has a constant factor approximation algorithm when each inactive social node requires only one active incoming neighboring node to become active (i.e., for all $j\in \mathcal{V}, \theta_j=1$). Therefore, we will analyze Problem \ref{problem} under the following assumption in this paper.

\begin{assumption} \label{assum:atleast_one_active_neighbor}
For all $j\in \mathcal{V},\theta_j=1$.
\end{assumption}

With regard to the physical nodes, we will consider the scenario where the total number of physical and social nodes are equal and each physical node covers exactly one social node (i.e., there is a bijective  mapping between the physical and social nodes). In the context of disaster recovery, this represents the case when a social group such as a household makes the decision to return or not after a disaster depending on whether an infrastructure component such as power connection at its home has been restored or not; thus, there is a one-to-one mapping between the household and the power connection at its home. Therefore, we make the following assumption (along with Assumption \ref{assum:atleast_one_active_neighbor}) in this paper.
\begin{assumption} \label{assum:no_secondary_edges}
$M=N$ and for all $l\in \mathcal{W},|\mathcal{Q}_l|=1$, with $\cup_{l \in \mathcal{W}}Q_l = \mathcal{V}$.
\end{assumption}

The following result shows that Problem \ref{problem} remains challenging \textit{even} under Assumptions \ref{assum:atleast_one_active_neighbor} and \ref{assum:no_secondary_edges}.  

\begin{proposition} \label{prop:NPhardness}
Problem \ref{problem} under Assumptions \ref{assum:atleast_one_active_neighbor} and \ref{assum:no_secondary_edges} is NP-hard.
\end{proposition}

The above result follows by noting that by choosing $K_p=M$ in the above problem, the physical nodes are removed from consideration, and we get back to the problem that is proved to be NP-hard by \cite{kempe2003maximizing} (see Theorem 2.4 of \cite{kempe2003maximizing}).

Since Problem \ref{problem} under Assumptions \ref{assum:atleast_one_active_neighbor} and \ref{assum:no_secondary_edges} is NP-hard, it is not possible to compute the optimal solution in polynomial-time, unless P = NP. Therefore, we characterize approximation algorithms for special cases of the problem in the next section. 

\section{Approximation algorithms} \label{sec:approx_algo}
We first provide the definition of \textit{reachability} in a network.
\begin{definition}[Reachability]
A social node $j\in \mathcal{V}$ is said to be reachable from a set $\mathcal{A}\subseteq\mathcal{V}$, if there exists a path in the graph $G=\{\mathcal{V},\mathcal{E}\}$ starting from a node $i\in \mathcal{A}$ and ending in node $j$.
\end{definition}

Note that in the above definition it is assumed that each node $j\in \mathcal{V}$ is reachable from itself. We now define $\sigma(\mathcal{A})$ and $\sigma_w(\mathcal{A})$.
\begin{definition}
Let $\mathcal{A}\subseteq \mathcal{V}$ be a set of social nodes and $\mathcal{B}$ be the set of all social nodes that are reachable from the set $\mathcal{A}$. Then we define $\sigma(\mathcal{A}) \triangleq |\mathcal{B}|$ and  $\sigma_w(\mathcal{A})\triangleq \sum_{j\in\mathcal{B}}w_j$.
\end{definition}

Note that $\sigma_w(\mathcal{A})$ can be exactly computed in polynomial-time (e.g., using Depth First Search (DFS) or Breadth First Search (BFS) \cite{cormen2009introduction}). 
We will now present some useful properties of $\sigma_w(\mathcal{A})$. We first present the following definitions.  
\begin{definition}
A set function $f$ is said to be monotone if $f(\mathcal{A}\cup \{j\})\ge f(\mathcal{A}), \forall j, \mathcal{A}$.
\end{definition}

\begin{definition}[Submodular function \cite{nemhauser1978analysis}]
A set function $f$ is said to be submodular if $f(\mathcal{A}\cup \{j\})-f(\mathcal{A})\ge f(\mathcal{B}\cup \{j\})-f(\mathcal{B}), \forall j, \mathcal{A}, \mathcal{B}$, when $\mathcal{A}\subseteq \mathcal{B}$ and $j\notin\mathcal{B} $.
\end{definition}

We have the following result.

\begin{lemma}
The function $\sigma_w$ is monotone and  submodular.
\end{lemma}

\begin{IEEEproof}
The proof of monotonicity comes directly from the fact that for any $\mathcal{A}\subseteq \mathcal{V}$ and $j\in \mathcal{V}$, the total weight of the social nodes that are reachable from the set $\mathcal{A}\cup \{j\}$ is at least equal to the total weight of the social nodes that are reachable from the set $\mathcal{A}$. 

We now prove submodularity. Consider a social node $j\in \mathcal{V}\setminus \mathcal{B}$ and two sets of social nodes $\mathcal{A}$ and $\mathcal{B}$ such that $\mathcal{A}\subseteq \mathcal{B}\subseteq\mathcal{V}$. Let $\mathcal{C}$ be the set of social nodes that are reachable from social node $j$ but not reachable from any node of the set $\mathcal{A}$. Let $\mathcal{D}$ be the set of social nodes that are reachable from social node $j$ but not reachable from any node of the set $\mathcal{B}$. Then, $\mathcal{D}\subseteq \mathcal{C}$ because $\mathcal{A}\subseteq \mathcal{B}$. Then, $\sigma_w(\mathcal{A}\cup \{j\})-\sigma_w(\mathcal{A})=\sum_{i\in \mathcal{C}}w_i\ge \sum_{i\in \mathcal{D}}w_i=\sigma_w(\mathcal{B}\cup \{j\})-\sigma_w(\mathcal{B})$. Thus, the result follows.
\end{IEEEproof}

We now present a greedy algorithm (see Algorithm \ref{alg:greedy_algo}) that we will use in the subsequent analysis.
\begin{algorithm}   \caption{Greedy selection of social nodes} \label{alg:greedy_algo}
Suppose Assumptions \ref{assum:atleast_one_active_neighbor} and \ref{assum:no_secondary_edges} hold. 
Set iteration $i=0$ and $\mathcal{A}_0=\emptyset$. 
  \begin{algorithmic}[1] 
    \State For $i=1$ to $K_{s}$, do the following.
    \begin{itemize}
        \item Let $j\in \mathcal{V}\setminus \mathcal{A}_{i-1}$ be a social node such that $ j\in\argmax_{c\in \mathcal{V}\setminus \mathcal{A}_{i-1}} \sigma_w(\mathcal{A}_{i-1}\cup \{c\})-\sigma_w(\mathcal{A}_{i-1})$ (breaking ties between social nodes by choosing the social node with largest weight). Define $\mathcal{A}_{i}=\mathcal{A}_{i-1}\cup \{j\}$.
    \end{itemize}
    \State Output $\mathcal{A}_{K_s}$ and $\sigma_w\left(\mathcal{A}_{K_s}\right)$.
  \end{algorithmic} 
\end{algorithm}
Note that Algorithm \ref{alg:greedy_algo} has polynomial-time complexity because Step 1 involves $K_s$ iterations where each iteration involves performing a max operation over an $O(N)$ array and $\sigma_w$ function can be computed in polynomial-time as argued earlier. We will use the following result in our analysis later (this result is inspired from Theorem 3 of \cite{lu2011approximation}; \cite{lu2011approximation} did not consider weighted social nodes, therefore we state the following result for the sake of completeness).

\begin{lemma} \label{lem:approx_nemhauser}
Let there be a graph $G=\{\mathcal{V},\mathcal{E}\}$ with $M (\geq 1)$ social nodes and a set $\mathcal{W}$ of $M$ physical nodes. Then, Algorithm \ref{alg:greedy_algo} is a $\frac{e}{e-1}$ approximation algorithm for Problem \ref{problem} when $K_p=M$, where $e$ is the base of the natural logarithm.
\end{lemma}

The proof of this result comes from the fact that when $K_p=M$, all physical nodes can be opened, and that $\sigma_w$ is a monotone submodular function \cite{nemhauser1978analysis}.

We now present an approximation algorithm (see Algorithm \ref{alg:approx_unweightednodes}) for Problem \ref{problem} under Assumptions \ref{assum:atleast_one_active_neighbor} and \ref{assum:no_secondary_edges}, for general $K_p\le M$.

\begin{algorithm}   \caption{Selection of seed nodes and opening of physical nodes} \label{alg:approx_unweightednodes}

%Let there be $M (\geq 1)$ social groups. 
  \begin{algorithmic}[1] 
    \State Run Algorithm \ref{alg:greedy_algo} to obtain the set $\mathcal{A}_{K_s}$.
    \State Suppose Assumptions \ref{assum:atleast_one_active_neighbor} and \ref{assum:no_secondary_edges} hold (and $K_p\le M$). Consider the following cases.
    \begin{enumerate}
        \item If $K_{s} > K_{p}$, then we select $K_p$ social nodes with the largest weights among all the nodes in $\mathcal{V}$ as the seed nodes and open the corresponding physical nodes of those seed nodes.
        \item If $K_{p} \ge \sigma \left(\mathcal{A}_{K_s}\right) \ge K_s$, then we first select the nodes in the set $\mathcal{A}_{K_s}$ as the seed nodes. After this, we open the physical nodes corresponding to all the social nodes that are reachable from the set $\mathcal{A}_{K_s}$.
        \item If $K_s \le K_{p} < \sigma \left(\mathcal{A}_{K_s}\right)$, we first select the set $\mathcal{A}_{K_s}$ as the seed set and open their corresponding physical nodes. We color all the social nodes white, except the seed nodes which are colored black. After this, we simulate the following process to open $K_p-K_s$ additional physical nodes. At every time-step in the simulation, we select the node $j$ with the largest weight among all the white nodes that have at least one black incoming neighbor, color node $j$ black and open the physical node of node $j$, until $K_p$ physical nodes are open. 
    \end{enumerate}
  \end{algorithmic} 
\end{algorithm}

Note that Algorithm \ref{alg:approx_unweightednodes} has polynomial-time complexity because of the following arguments. The first step of the algorithm has polynomial-time complexity because Algorithm \ref{alg:greedy_algo} is a polynomial-time algorithm. We now focus on the complexity of Step 2 of Algorithm \ref{alg:approx_unweightednodes}. The complexity of case 1 is polynomial-time as $O(K_p)$ operations are required; note that $K_p=O(N)$ as $K_p \le M=N$. Also, note that $\sigma (\mathcal{A}_{K_s})$ can be computed in polynomial-time as argued earlier. Therefore, case 2 has polynomial-time complexity because all the reachable nodes from $\mathcal{A}_{K_s}$ can be identified in polynomial-time by DFS \cite{cormen2009introduction}. Case 3 has polynomial-time complexity because after setting the seed set, the simulation takes $O(K_p-K_s)$ operations. We now prove that Algorithm \ref{alg:approx_unweightednodes} is an approximation algorithm.

\begin{theorem} \label{thm:scc_socialgroups_greedy}
Suppose Assumptions \ref{assum:atleast_one_active_neighbor} and \ref{assum:no_secondary_edges} hold. Let there be a graph $G=\{\mathcal{V},\mathcal{E}\}$ with $M (\geq 1)$ social nodes and a set $\mathcal{W}$ of $M$ physical nodes. Let the largest and the smallest weights of the social nodes be $w^{max}$ and $w^{min}$, respectively. Then, Algorithm \ref{alg:approx_unweightednodes} is a $\max\{\frac{e}{e-1},\frac{w^{max}}{w^{min}}\}$ approximation algorithm.
\end{theorem}
\begin{IEEEproof}
Suppose $K_{s} > K_{p}$. Then, Algorithm \ref{alg:approx_unweightednodes} gives the optimal solution because the maximum number of social nodes that can eventually be activated is equal to $K_p$, and $K_p$ social nodes with the largest weights in the set $\mathcal{V}$ are activated by the algorithm. 

We now consider the case when $K_{s} \le K_{p}$. Note that $\sigma(\mathcal{A}_{K_s})\ge K_s$ because each social node is reachable from itself. Thus, there are two subcases when $K_{s} \le K_{p}$: (i) $K_{p} \ge \sigma \left(\mathcal{A}_{K_s}\right)$ and (ii) $K_{p} < \sigma \left(\mathcal{A}_{K_s}\right)$. Suppose $K_{p} \ge \sigma \left(\mathcal{A}_{K_s}\right) \ge K_s$. Then, all the reachable nodes from the seed set $\mathcal{A}_{K_s}$ can eventually be activated. Denote $\mathcal{A}^*\in\argmax_{|\mathcal{A}|\le K_s}\sigma_w(\mathcal{A})$ as the optimal seed set when $K_p=M$, and Assumptions \ref{assum:atleast_one_active_neighbor} and \ref{assum:no_secondary_edges} hold. Let $C$ be the optimal value of Problem \ref{problem} under Assumptions \ref{assum:atleast_one_active_neighbor} and \ref{assum:no_secondary_edges}. Then,
\begin{equation*}
 \frac{C}{\sigma_w(\mathcal{A}_{K_s})}\le \frac{\sigma_w(\mathcal{A}^*)}{\sigma_w(\mathcal{A}_{K_s})} \le \frac{e}{e-1}, \label{eq:Kp_greaterthan_sigma}
\end{equation*}
where the left most inequality comes from the fact that $C \le \sigma_w(\mathcal{A}^*)$ because $\sigma_w(\mathcal{A}^*)$ is the optimal value when Assumptions \ref{assum:atleast_one_active_neighbor} and \ref{assum:no_secondary_edges} hold but there is no constraint on the total number of physical nodes that can be opened, and the last inequality comes from Lemma \ref{lem:approx_nemhauser}.  

Suppose $K_s \le K_{p} < \sigma \left(\mathcal{A}_{K_s}\right)$. Note that the number of social nodes that eventually get activated by Algorithm \ref{alg:approx_unweightednodes} is equal to the number of black colored nodes, which is equal to $K_p$. 
Let $x$ be the number of social nodes that eventually get activated by the optimal solution. Denote the optimal value as $C$ and let the total weight of the social nodes that are eventually activated by Algorithm \ref{alg:approx_unweightednodes} be $C'$. Then, $C\le x w^{max}$ and $C'\ge K_p w^{min}$ from the definitions of $w^{max}$ and $w^{min}$, respectively. Thus, $\frac{C}{C'}\le \frac{x w^{max}}{K_p w^{min}}\le \frac{K_p w^{max}}{K_p w^{min}}=\frac{w^{max}}{w^{min}}$ as $x\le K_p$.

Combining the bounds from each of the above three cases we see that Algorithm \ref{alg:approx_unweightednodes} is a $\max\{\frac{e}{e-1},\frac{w^{max}}{w^{min}}\}$ approximation algorithm.
\end{IEEEproof}
\begin{figure}[ht]
	\begin{center}
		\includegraphics[scale=0.4]{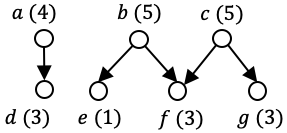}
	\end{center}	
	\caption{Graph for illustrating Algorithm \ref{alg:approx_unweightednodes}.}
	\label{fig:general-graph}
\end{figure}

We now provide an example to illustrate Algorithm \ref{alg:approx_unweightednodes}.

\begin{example}
Consider a social network as shown in Figure \ref{fig:general-graph} with the corresponding weights shown in the parentheses. Suppose that Assumptions \ref{assum:atleast_one_active_neighbor} and \ref{assum:no_secondary_edges} hold. Thus, there is a physical node corresponding to each social node but physical nodes are not shown in the figure for simplicity. Suppose $K_s=2$ and $K_p=4$. Then, nodes $c$ and $a$ are selected as the seed nodes when $\mathcal{A}_{K_s}$ is determined by Algorithm \ref{alg:greedy_algo}. Thus, $\sigma(\mathcal{A}_{K_s})=5$. Note that $\sigma(\mathcal{A}_{K_s})=5>4=K_p$. Thus, $K_s \le K_{p} < \sigma \left(\mathcal{A}_{K_s}\right)$ holds and the nodes $\{a,c,d,f\}$ are activated when Algorithm \ref{alg:approx_unweightednodes} is applied with the total weight of the activated nodes being equal to 15 (note that any pair of nodes from the set $\{d,f,g\}$ could have been activated after the selection of seed nodes). Note that it is optimal to select nodes $b$ and $c$ as the seed nodes and open the physical nodes corresponding to $b,c,f$ and $g$ with the total weight of the activated nodes being equal to 16. Note that although Algorithm \ref{alg:approx_unweightednodes} is not optimal in this example, it is indeed a $\max\{\frac{e}{e-1},\frac{w^{max}}{w^{min}}\}=\max\{\frac{e}{e-1},5\}=5$ approximation algorithm as proved in Theorem \ref{thm:scc_socialgroups_greedy}.
\end{example}

We now focus on another instance of Problem \ref{problem} under Assumptions \ref{assum:atleast_one_active_neighbor} and \ref{assum:no_secondary_edges}. Consider the following assumption.
\begin{assumption} \label{assum:bipartite}
Suppose the graph $G=\{\mathcal{V},\mathcal{E}\}$ is a directed bipartite graph consisting of two sets of social nodes $\mathcal{I}$ and $\mathcal{J}$ such that $\mathcal{I} \cup \mathcal{J}=\mathcal{V}$ and $\mathcal{I} \cap \mathcal{J}=\emptyset$. Also, suppose that every edge of the set $\mathcal{E}$ starts from a node of the set $\mathcal{I}$ and ends in a node of the set $\mathcal{J}$ such that there is at least one edge that starts from every node of the set $\mathcal{I}$ and there is at least one edge that ends in every node of the set $\mathcal{J}$. Suppose the weights of all the nodes of the set $\mathcal{I}$ lie in the interval $[\underline{w}_{\mathcal{I}}, \overline{w}_{\mathcal{I}}]$ and the weights of all the nodes of the set $\mathcal{J}$ lie in the interval $[\underline{w}_{\mathcal{J}}, \overline{w}_{\mathcal{J}}]$ such that $\underline{w}_{\mathcal{I}}\ge \overline{w}_{\mathcal{J}}$.
\end{assumption}

In the context of disaster recovery, Problem \ref{problem} under Assumptions \ref{assum:atleast_one_active_neighbor}, \ref{assum:no_secondary_edges} and \ref{assum:bipartite} represents the scenario where small communities (represented by the set $\mathcal{J}$) are socially influenced by larger communities (represented by the set $\mathcal{I}$). 

Note that Problem \ref{problem} under Assumptions \ref{assum:atleast_one_active_neighbor}, \ref{assum:no_secondary_edges} and \ref{assum:bipartite} is NP-hard
because by setting $K_p=M$, it is possible to open all physical nodes and we get back to the problem that has been proved to be NP-hard in Theorem 2.4 of \cite{kempe2003maximizing}. We now present the following result.

\begin{theorem} \label{thm:bipartite_graph}
Suppose Assumptions \ref{assum:atleast_one_active_neighbor}, \ref{assum:no_secondary_edges} and \ref{assum:bipartite} hold, i.e., there is a bipartite directed graph $G=\{\mathcal{V},\mathcal{E}\}$ with $M (\geq 1)$ social nodes such that $\mathcal{I} \cup \mathcal{J}=\mathcal{V}$ and $\mathcal{I} \cap \mathcal{J}=\emptyset$, along with a set $\mathcal{W}$ of $M$ physical nodes. Then, Algorithm \ref{alg:approx_unweightednodes} is a $\max\{\frac{e}{e-1},\frac{\overline{w}_{\mathcal{I}} \hspace{0.7mm} \overline{w}_{\mathcal{J}}}{\underline{w}_{\mathcal{I}} \hspace{0.7mm} \underline{w}_{\mathcal{J}}}\}$ approximation algorithm.
\end{theorem}

\begin{IEEEproof}
Note that the proofs for the cases when $K_s> K_p$ and $K_{p} \ge \sigma \left(\mathcal{A}_{K_s}\right) \ge K_s$ follow in the same way as in the proof of Theorem \ref{thm:scc_socialgroups_greedy}. Thus, we focus on the case when $K_s \le K_{p} < \sigma \left(\mathcal{A}_{K_s}\right)$. Suppose $K_s\le |\mathcal{I}|$. Then, $\mathcal{A}_{K_s}\subseteq \mathcal{I}$ because the weight of each node $i\in \mathcal{I}$ is larger than or equal to the weight of each node $j\in \mathcal{J}$ (as $\underline{w}_{\mathcal{I}}\ge \overline{w}_{\mathcal{J}}$), and for all nodes $j\in \mathcal{J}$, there is no node $j'\in \mathcal{V}$ such that $j$ is an incoming neighbor of $j'$. Denote the optimal value as $C$ and let the total weight of the social nodes that are eventually activated by Algorithm \ref{alg:approx_unweightednodes} be $C'$. Recall that Algorithm \ref{alg:approx_unweightednodes} eventually activates $K_p$ social nodes when $K_s \le K_{p} < \sigma \left(\mathcal{A}_{K_s}\right)$. Thus, $C\le K_s \overline{w}_{\mathcal{I}}+(K_p-K_s)\overline{w}_{\mathcal{J}}$ and $C'\ge K_s \underline{w}_{\mathcal{I}}+(K_p-K_s)\underline{w}_{\mathcal{J}}$ because it is not possible to activate more than $K_s$ nodes in the set $\mathcal{I}$ as no node in the set $\mathcal{I}$ has any incoming neighboring node and $\mathcal{A}_{K_s}\subseteq \mathcal{I}$. Note that 
\begin{align*}
    C'\overline{w}_{\mathcal{I}} \hspace{0.7mm} \overline{w}_{\mathcal{J}}
    &\ge  K_s \underline{w}_{\mathcal{I}}\overline{w}_{\mathcal{I}} \hspace{0.7mm} \overline{w}_{\mathcal{J}}+(K_p-K_s)\underline{w}_{\mathcal{J}}\overline{w}_{\mathcal{I}} \hspace{0.7mm} \overline{w}_{\mathcal{J}} \\
    &\ge  K_s \underline{w}_{\mathcal{I}}\overline{w}_{\mathcal{I}} \hspace{0.7mm} \underline{w}_{\mathcal{J}}+(K_p-K_s)\underline{w}_{\mathcal{J}}\underline{w}_{\mathcal{I}} \hspace{0.7mm} \overline{w}_{\mathcal{J}}\\
  &\ge C\underline{w}_{\mathcal{I}} \hspace{0.7mm} \underline{w}_{\mathcal{J}},
\end{align*} 
because $\overline{w}_{\mathcal{J}}\ge \underline{w}_{\mathcal{J}}$ and  $\overline{w}_{\mathcal{I}}\ge \underline{w}_{\mathcal{I}}$. Therefore, $
    \frac{C}{C'}\le \frac{\overline{w}_{\mathcal{I}} \hspace{0.7mm} \overline{w}_{\mathcal{J}}}{\underline{w}_{\mathcal{I}} \hspace{0.7mm} \underline{w}_{\mathcal{J}}}$.

Suppose $K_s > |\mathcal{I}|$. Then, $\mathcal{I} \subset \mathcal{A}_{K_s}$ because of the same argument as given for $\mathcal{A}_{K_s}\subseteq \mathcal{I}$ when $K_s\le |\mathcal{I}|$. Thus, $\mathcal{A}_{K_s}$ contains the top $K_s$ social nodes with largest weight in the set $\mathcal{V}$ because in each iteration $i>|\mathcal{I}|$ of Algorithm \ref{alg:greedy_algo}, $\sigma_w(\mathcal{A}_{i-1}\cup j)-\sigma_w(\mathcal{A}_{i-1})=\sum_{k\in \mathcal{V}}w_k-\sum_{k\in \mathcal{V}}w_k=  0, \forall j \in \mathcal{V}\setminus \mathcal{A}_{i-1}$ (as for all $i>|\mathcal{I}|$,  $\mathcal{I}\subseteq \mathcal{A}_{i-1}$ and each node $j \in \mathcal{V}\setminus \mathcal{A}_{i-1}$ has an incoming neighbor in the set $\mathcal{I}$) and ties are resolved in Algorithm \ref{alg:greedy_algo} by choosing the social node with largest weight. Therefore, Algorithm \ref{alg:approx_unweightednodes} gives the optimal solution because after opening the physical nodes corresponding to the seed nodes in Algorithm \ref{alg:approx_unweightednodes}, physical nodes corresponding to the top $K_p-K_s$ social nodes with largest weight in the set $\mathcal{V}\setminus \mathcal{A}_{K_s}$ are opened (since $\mathcal{V}\setminus \mathcal{A}_{K_s} \subset \mathcal{J}$).    

Thus, the approximation ratio of Algorithm \ref{alg:approx_unweightednodes} is $\max\{\frac{e}{e-1},\frac{\overline{w}_{\mathcal{I}} \hspace{0.7mm} \overline{w}_{\mathcal{J}}}{\underline{w}_{\mathcal{I}} \hspace{0.7mm} \underline{w}_{\mathcal{J}}}\}$. %Thus, the result follows.  
\end{IEEEproof}
\begin{remark}
Note that the approximation ratio characterized by Algorithm \ref{alg:approx_unweightednodes} in Theorem \ref{thm:bipartite_graph} is less than or equal to the one characterized in Theorem \ref{thm:scc_socialgroups_greedy} for Problem \ref{problem} under Assumptions \ref{assum:atleast_one_active_neighbor}, \ref{assum:no_secondary_edges} and \ref{assum:bipartite} because the largest and the smallest weights among all the social nodes are $\overline{w}_{\mathcal{I}}$ and $ \underline{w}_{\mathcal{J}}$, respectively, and $\frac{\overline{w}_{\mathcal{I}} \hspace{0.7mm} \overline{w}_{\mathcal{J}}}{\underline{w}_{\mathcal{I}} \hspace{0.7mm} \underline{w}_{\mathcal{J}}} \le \frac{\overline{w}_{\mathcal{I}} }{ \underline{w}_{\mathcal{J}}}$ (as $\underline{w}_{\mathcal{I}}\ge \overline{w}_{\mathcal{J}}$).
\end{remark}

Suppose $\underline{w}_{\mathcal{I}}=\overline{w}_{\mathcal{I}}$ and $\underline{w}_{\mathcal{J}}=\overline{w}_{\mathcal{J}}$. Then, the following result holds by Theorem \ref{thm:bipartite_graph} because $\max\{\frac{e}{e-1},\frac{\overline{w}_{\mathcal{I}} \hspace{0.7mm} \overline{w}_{\mathcal{J}}}{\underline{w}_{\mathcal{I}} \hspace{0.7mm} \underline{w}_{\mathcal{J}}}\}=\max\{\frac{e}{e-1},1\}=\frac{e}{e-1}$.
\begin{corollary} 
Suppose $\underline{w}_{\mathcal{I}}=\overline{w}_{\mathcal{I}}$, $\underline{w}_{\mathcal{J}}=\overline{w}_{\mathcal{J}}$ and Assumptions \ref{assum:atleast_one_active_neighbor}, \ref{assum:no_secondary_edges} and \ref{assum:bipartite} hold, i.e., there is a bipartite directed graph $G=\{\mathcal{V},\mathcal{E}\}$ with $M (\geq 1)$ social nodes such that $\mathcal{I} \cup \mathcal{J}=\mathcal{V}$ and $\mathcal{I} \cap \mathcal{J}=\emptyset$, along with a set $\mathcal{W}$ of $M$ physical nodes. Then, Algorithm \ref{alg:approx_unweightednodes} is a $\frac{e}{e-1}$ approximation algorithm.
\end{corollary}

\subsection{Evaluation of Algorithm \ref{alg:approx_unweightednodes}}
We now evaluate the performance of the approximation algorithm that we characterized (Algorithm \ref{alg:approx_unweightednodes}) with a \textit{brute-force method}. In the brute-force method, all the possible combinations of $K_s$ seed nodes and $K_p$ open physical nodes are enumerated to find an optimal solution. Consider the social network as shown in Figure \ref{fig:general-graph} that satisfies Assumptions \ref{assum:atleast_one_active_neighbor}, \ref{assum:no_secondary_edges} and \ref{assum:bipartite} such that $N=M=7$, $|\mathcal{I}|=3$, $\overline{w}_{\mathcal{I}}=5,\underline{w}_{\mathcal{I}}=4$, $\overline{w}_{\mathcal{J}}=3$ and $ \underline{w}_{\mathcal{J}}=1$. Suppose $K_s=2$ and $K_p=4$. 
Consider the first row of Table \ref{tab:enumeration}. Then, the second column of that row shows the computation time (in seconds) for the brute-force method, the third column shows the computation time (in seconds) for Algorithm \ref{alg:approx_unweightednodes} and the fourth column shows the ratio of the optimal value (computed by the brute-force method) with respect to the value computed by Algorithm \ref{alg:approx_unweightednodes}. In the subsequent rows, we increase $N, K_s$ and $K_p$ such that additional social nodes (and thus physical nodes) are added in the set $\mathcal{J}$ such that Assumptions \ref{assum:atleast_one_active_neighbor}, \ref{assum:no_secondary_edges} and \ref{assum:bipartite} hold. We can see that the computation time for the brute-force method increases rapidly with the size of the problem but the increase in the computation time for Algorithm \ref{alg:approx_unweightednodes} is much slower with the problem size. Also, note that $\max\{\frac{e}{e-1},\frac{\overline{w}_{\mathcal{I}} \hspace{0.7mm} \overline{w}_{\mathcal{J}}}{\underline{w}_{\mathcal{I}} \hspace{0.7mm} \underline{w}_{\mathcal{J}}}\}=\max\{\frac{e}{e-1},\frac{15}{4}\}=3.75$ for all the considered instances in this example. Thus, the ratio of the optimal value to the value computed by Algorithm \ref{alg:approx_unweightednodes} would not exceed 3.75 by Theorem \ref{thm:bipartite_graph}. Therefore, brute-force method is not an efficient method for solving the problem, illustrating the benefit of our approximation algorithms.

\begin{table}[h]
	\caption{Results when the problem parameters are varied}
	\label{tab:enumeration}
	\begin{center}
		\begin{tabular}{|cccc|}
			\hline
			Parameters& Brute-force &Algorithm \ref{alg:approx_unweightednodes} & Ratio of optimal  \\ 
			$(N,K_s,K_p)$& time (s) & time (s) & to approx. value \\ \hline
			$7,2,4$&0.01& 0.004 &1.07\\
			$9,3,5$&0.07& 0.005&1.00 \\
			$11,4,6$&0.65& 0.005 &1.15\\ 
			$13,5,7$&9.85& 0.005 &1.20\\ 
			$15,6,8$&164.14& 0.005 &1.25\\ \hline
		\end{tabular}
	\end{center}
\end{table}

All the instances of Problem \ref{problem} that we have considered until now are NP-hard (and thus it is not possible to efficiently compute the optimal solution). We present a special instance of Problem \ref{problem} under Assumptions \ref{assum:atleast_one_active_neighbor} and \ref{assum:no_secondary_edges} in the next section that can be optimally solved in polynomial-time.

%I can write the section as instances that can be exactly solved in polynomial-time
\section{Optimal algorithm when the social network is a disjoint union of out-trees}
We start by defining an \textit{out-tree} as follows.
\begin{definition}
An \textit{out-tree} is a directed rooted tree with all the edges pointed away from the root node. 
\end{definition}

We make the following assumption (along with Assumptions \ref{assum:atleast_one_active_neighbor} and \ref{assum:no_secondary_edges}) in this section.
\begin{assumption} \label{assum:forest_out-tree}
$G=\{\mathcal{V},\mathcal{E}\}$ is a set of disjoint out-trees (i.e., a forest of out-trees).
\end{assumption}

Note that there are several studies that have observed the presence of tree-type hierarchical structures in social networks \cite{gilbert2011communities,martin1998structures,wang2013exponential,shumate2010exponential,deepa2019centrality}. For instance, a social network in the form of a single out-tree could represent multiple social groups that have a hierarchy of power or influence between them, with the social group corresponding to the root node being the most influential \cite{martin1998structures}. In addition, directed star graphs\footnote{A star graph is a tree that has a single node with more than one neighbors.} such as out-stars are frequently used in social network analysis to represent bi-level hierarchies between social groups \cite{wang2013exponential,shumate2010exponential}. Also, the paper \cite{deepa2019centrality} collected social network information of households in a village and found that the social network forms an out-star during crisis periods (such as disasters) where the root node denotes an influential household of the village.    

\begin{figure}[ht]
	\begin{center}
		\includegraphics[scale=0.4]{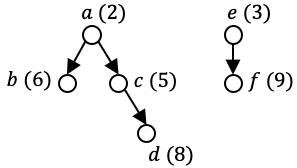}
	\end{center}	
	\caption{Graph for illustrating Example \ref{exmp:forest_out-tree_non-trivial}.}
	\label{fig:non-trivial-forest_out-tree}
\end{figure}

We first discuss how the presence of physical nodes and weighted social nodes make our problem more challenging and interesting than in the case without physical nodes, even under Assumption \ref{assum:forest_out-tree}. Consider the instance of Problem \ref{problem} under Assumptions \ref{assum:atleast_one_active_neighbor}, \ref{assum:no_secondary_edges} and \ref{assum:forest_out-tree} when $K_p=M$, i.e., it is possible to open all physical nodes. Then, the optimal solution for this case can be easily computed as follows: select the $K_s$ out-trees that have the largest total weight of social nodes, set the root nodes of the selected out-trees as the seed nodes and open all the $M$ physical nodes. 
Similarly, consider another instance of Problem \ref{problem} under Assumptions \ref{assum:atleast_one_active_neighbor}, \ref{assum:no_secondary_edges} and \ref{assum:forest_out-tree} when $w_j=w, \forall j \in \mathcal{V}$, i.e., the weights of all the social nodes are the same (for general $K_p\le M$). Then, the optimal solution can be easily computed by repeating the following procedure until $K_s$ seed nodes are selected or $K_p$ physical nodes are open: select an out-tree with the largest number of nodes among the set of out-trees that have not been selected before, set the root node $j$ of the out-tree as the seed node, and open the physical nodes corresponding to an arbitrary out-tree of size $l$ that is rooted at node $j$, where $l$ is the minimum of the size of the selected out-tree and the remaining number of physical nodes that can be opened.
However, the optimal strategy is not at all obvious when the social nodes have heterogeneous weights and there is a constraint on the number of physical nodes that can be opened as argued in the next example. 
\begin{example} \label{exmp:forest_out-tree_non-trivial}
Consider a social network as shown in Figure \ref{fig:non-trivial-forest_out-tree} with the corresponding weights shown in parentheses. Suppose that Assumptions \ref{assum:atleast_one_active_neighbor}, \ref{assum:no_secondary_edges} and \ref{assum:forest_out-tree} hold such that $N=M=6$. There is a physical node corresponding to each social node (physical nodes are not shown in the figure for simplicity). Suppose $K_s=2$ and $K_p=3$. Since $K_p=3<6=N$ it is not possible to open all the physical nodes and thus the above algorithm that allowed all physical nodes to open cannot be used to find the optimal solution. Also, if the algorithm that assumed homogeneous weights for the social nodes is used, then node $a$ would be set as a seed node and physical nodes corresponding to an out-tree that is rooted at node $a$ but has three social nodes would be opened. However, the aforementioned solution would not be optimal because the optimal solution is to select nodes $c$ and $f$ as the seed nodes and open the physical nodes corresponding to the social nodes $c,d$ and $f$; the total weight of the activated nodes in the optimal solution is 22. 
\end{example}

Note that there may even be social networks where multiple seed nodes are required in one out-tree in the optimal solution (i.e., non-contiguous portions of the out-tree need to be activated), which is not the case in situations without physical nodes. Thus, we will present an algorithm (Algorithm \ref{alg:DP_forest_outtree}) in the following discussion that is optimal for Problem \ref{problem} under Assumptions \ref{assum:atleast_one_active_neighbor}, \ref{assum:no_secondary_edges} and \ref{assum:forest_out-tree} (even when the weights are heterogeneous across the social nodes and it may not be possible to open all the physical nodes).    

\begin{figure}[ht]
	\begin{center}
		\includegraphics[scale=0.4]{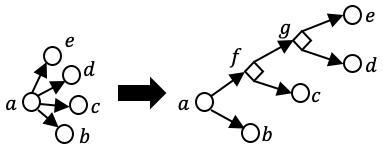}
	\end{center}	
	\caption{Converting an out-tree (left) to another out-tree (right) where each node has at most two outgoing neighbors (the diamonds represent dummy nodes).}
	\label{fig:out-tree_dummy_nodes_addition}
\end{figure}

 In the first step of Algorithm \ref{alg:DP_forest_outtree}, $G$ is modified to another forest $\overline{G}$ by running Algorithm \ref{alg:outtree_dummy_addition} on each out-tree of $G$. In Algorithm \ref{alg:outtree_dummy_addition}, dummy nodes with zero weights are added so that each node in the modified network has at most two outgoing neighbors;
the condition that each node has at most two outgoing nodes ensures that Algorithm \ref{alg:DP_forest_outtree} has polynomial-time complexity (we discuss this later in Remark \ref{rem:complexity_forest_out-tree}). Figure \ref{fig:out-tree_dummy_nodes_addition} shows a forest $G$ containing a single out-tree on the left-hand side (LHS) where node $a$ has more than two outgoing neighbors; on the right-hand side (RHS) is an out-tree $\overline{G}$ that is generated by adding dummy nodes $f$ and $g$ so that each node has at most two outgoing neighbors. In the second step of Algorithm \ref{alg:DP_forest_outtree}, $\overline{G}$ is modified to an out-tree $G'$ through the addition of dummy nodes. Finally, Algorithm \ref{alg:DP_out-tree} is run on $G'$ to obtain an optimal solution in the last step of Algorithm \ref{alg:DP_forest_outtree}. Note that Algorithm \ref{alg:DP_out-tree} is a Dynamic Programming algorithm that first computes the optimal values for the out-trees rooted at the outgoing neighbors of each node $v $ in $G'$ before computing the optimal value for the out-tree rooted at $v$; we  provide more details on the parameters that are computed in Algorithm \ref{alg:DP_out-tree} later. Note that while choosing a solution in the last step of Algorithm \ref{alg:DP_forest_outtree}, a dummy node is not set as a seed node. Also, physical nodes are not mapped to the dummy nodes in the aforementioned steps and thus the only necessary condition for an inactive dummy node to become active is that at least one of its incoming neighboring nodes should be active.

\begin{algorithm} \caption{Optimal algorithm when the social network is a forest of out-trees} \label{alg:DP_forest_outtree}
Consider a social network $G$ that is a forest of out-trees. 
\begin{algorithmic}[1] 
\State Run Algorithm \ref{alg:outtree_dummy_addition} on each of the out-trees of forest $G$ to obtain a modified forest $\overline{G}$.
\State If there is a single out-tree in $\overline{G}$, then set $G'=\overline{G}$ and proceed to the next step. Otherwise, construct an out-tree $G'$ as follows. Construct a dummy node $i$ with zero weight. Let $\mathcal{L}$ be the set containing all the root nodes of the out-trees in $\overline{G}$. 
% Order the nodes in $\mathcal{L}$ with the left most node to $i$ being $i_1$, the second left most node to $i$ being $i_2$ and so on. 
Set node $j=i$ and let $x$ be the number of out-trees in $\overline{G}$. Then, repeat the following until the termination criterion is reached. 
    \begin{itemize}
        \item Stop if $x<2$. If $x=2$, construct edges starting from node $j$ and ending in all the nodes in the set $\mathcal{L}$. Otherwise, construct an edge starting from node $j$ and ending in an arbitrary node $m$ in the set $\mathcal{L}$. Then, remove node $m$ from the set $\mathcal{L}$. Also, construct a dummy node $l$ with zero weight and an edge starting from node $j$ and ending in node $l$. Set $j=l$ and $x=x-1$. 
    \end{itemize}
      \State 
Run Algorithm \ref{alg:DP_out-tree} on $G'$ to find the seed nodes and the physical nodes to open.
\end{algorithmic}
\end{algorithm}

\begin{algorithm}   \caption{Addition of dummy nodes to an out-tree } \label{alg:outtree_dummy_addition}
Consider an out-tree $G''=\{\mathcal{V}'',\mathcal{E}''\}$. 
%Let there be $M (\geq 1)$ social groups. 
  \begin{algorithmic}[1] 
    \State Repeat the following until the termination criterion is reached.
    \begin{itemize}
        \item Stop if there is no node in $\mathcal{V}''$ that has more than two outgoing neighbors. Otherwise, arbitrarily select a node $i\in \mathcal{V}''$ that has more than two outgoing neighbors. 
        % Order the incoming neighbors of $i$ with the left most neighbor of $i$ being $i_1$, the second left most neighbor of $i$ being $i_2$ and so on. 
        Let $\mathcal{L}$ be the set containing all the outgoing neighbors of $i$. Remove all the edges between node $i$ and the nodes in  $\mathcal{L}$, and set node $j=i$ and $x=|\mathcal{L}|$. Then, repeat the following until the termination criterion is reached. 
    \begin{itemize}
        \item Stop if $x<2$. If $x=2$, construct edges starting from node $j$ and ending in all the nodes in the set $\mathcal{L}$. Otherwise, construct an edge starting from node $j$ and ending in an arbitrary node $m$ in the set $\mathcal{L}$. Then, remove node $m$ from the set $\mathcal{L}$. Also, construct a dummy node $l$ with zero weight and an edge starting from node $j$ and ending in node $l$. Then, set $j=l$ and $x=x-1$. 
    \end{itemize}
    \end{itemize}
    %   \State Denote the social network constructed above by $\overline{G}$. 
  \end{algorithmic}
\end{algorithm}

We now present the conditions that are used in Algorithm \ref{alg:DP_out-tree}. Consider an out-tree $G'=\{\mathcal{V}',\mathcal{E}'\}$, where each node has at most two outgoing neighbors. Denote the root node of $\mathcal{V}'$ as $r$. Let $f_v(k,l)$ be the optimal value of the total weight of eventually activated social nodes for the out-tree rooted at node $v\in \mathcal{V}'$ when there are at most $k$ seed nodes and at most $l$ open physical nodes in the out-tree rooted at node $v$. 
 Let $\overline{f}_v(k,l)$ (resp. $\underline{f}_v(k,l)$) be the optimal value of the total weight of eventually activated social nodes for the out-tree rooted at node $v\in \mathcal{V}'$ when there are at most $k$ seed nodes and at most $l$ open physical nodes in the out-tree rooted at $v$ such that node $v$ is active (resp. inactive). Note that if $v$ is a non-dummy node, then $\overline{f}_v(k,l)$ is only defined when  $l\ge 1$; if $v$ is a dummy node, then $\overline{f}_v(k,l)$ is only defined when $v\neq r$,\footnote{Recall that the root node can be a dummy node because we start by constructing a dummy node in Step 2 of Algorithm \ref{alg:DP_forest_outtree} if $\overline{G}$ contains more than one out-tree.} $ k \le K_s-1$ and $ l \le K_p-1$ (because the predecessor node of a dummy node needs to be active in order for the dummy node to be active). 
% Denote the predecessor node of each node $v\in \mathcal{V}'$ as $v_p$. 
Let $\overline{f}^a_v(k,l)$ be the optimal value of the total weight of eventually activated social nodes for the out-tree rooted at node $v\in \mathcal{V}'$ when there are at most $k$ seed nodes and at most $l$ open physical nodes in the out-tree rooted at $v$ such that node $v$ is active but is not a seed node. Note that if $v$ is a non-dummy node, then $\overline{f}^a_v(k,l)$ is only defined when $  k \le K_s-1$ and $1 \le l\le K_p-1$; if $v$ is a dummy node, then $\overline{f}^a_v(k,l)$ is only defined when $v\neq r$, $  k \le K_s-1$ and $l \le K_p-1$ (as the predecessor node of $v$ should also be active). Let $\overline{f}^b_v(k,l)$ be the optimal value of the total weight of eventually activated social nodes for the out-tree rooted at node $v\in \mathcal{V}'$ when there are at most $k$ seed nodes and at most $l$ open physical nodes in the out-tree rooted at $v$ such that node $v$ is active along with being a seed node; note that $\overline{f}^b_v(k,l)$ is only defined when $k\ge 1$ and $l\ge 1$, and only for non-dummy nodes because a dummy node cannot be set as a seed node. Let $\mathcal{F}_v(k,l)$ be a set of seed nodes and  $\mathcal{H}_v(k,l)$ be a set of physical nodes that are opened in the out-tree rooted at $v$ to obtain $f_v(k,l)$. Also, let $\overline{\mathcal{F}}_v(k,l)$ (resp. $ \underline{\mathcal{F}}_v(k,l)$) be a set of seed nodes and $\overline{\mathcal{H}}_v(k,l)$ (resp. $\underline{\mathcal{H}}_v(k,l)$) be a set of physical nodes that are opened in the out-tree rooted at $v$ to obtain $\overline{f}_v(k,l)$ (resp. $\underline{f}_v(k,l)$). Finally, let $\overline{\mathcal{F}}^a_v(k,l)$ (resp. $ \overline{\mathcal{F}}^b_v(k,l)$) be a set of seed nodes and $\overline{\mathcal{H}}^a_v(k,l)$ (resp. $\overline{\mathcal{H}}^b_v(k,l)$) be a set of physical nodes that are opened in the out-tree rooted at  $v$ to obtain $\overline{f}^a_v(k,l)$ (resp. $\overline{f}^b_v(k,l)$).

Note that for each node $v\in \mathcal{V}'$, if  $\overline{f}_v(k,l)$ is defined and $\overline{f}_v(k,l)\ge \underline{f}_v(k,l)$, then
\begin{align}
f_v(k,l)=\overline{f}_v(k,l),  \mathcal{F}_v(k,l)=\overline{\mathcal{F}}_v(k,l),\mathcal{H}_v(k,l)=\overline{\mathcal{H}}_v(k,l); \label{eq:f_v(k,l)_non_dummy_kge1_lge1_outtree} 
\end{align}
otherwise,
\begin{align}
f_v(k,l)=\underline{f}_v(k,l),\mathcal{F}_v(k,l)=\underline{\mathcal{F}}_v(k,l),\mathcal{H}_v(k,l)=\underline{\mathcal{H}}_v(k,l). \label{eq:f_v(k,l)_non_dummy_outtree}
\end{align}

Also, if only $\overline{f}^a_v(k,l)$ is defined or if both $\overline{f}^a_v(k,l)$ and $\overline{f}^b_v(k,l)$ are defined but $\overline{f}^a_v(k,l)\ge \overline{f}^b_v(k,l)$, then
\begin{equation}
\overline{f}_v(k,l)=\overline{f}^a_v(k,l), \overline{\mathcal{F}}_v(k,l)=\overline{\mathcal{F}}^a_v(k,l) ,\overline{\mathcal{H}}_v(k,l)=\overline{\mathcal{H}}^a_v(k,l);\label{eq:f_overline_nonroot}
\end{equation}
otherwise,
\begin{equation}
\overline{f}_v(k,l)=\overline{f}^b_v(k,l), \overline{\mathcal{F}}_v(k,l)=\overline{\mathcal{F}}^b_v(k,l),\overline{\mathcal{H}}_v(k,l)=\overline{\mathcal{H}}^b_v(k,l). \label{eq:f_overline_othercases}
\end{equation}

%The optimal value is $\max_v f_v(K_s,K_p)$. 

For each node $v\in \mathcal{V}'$, if $k=0$ or $l=0$,
\begin{equation}
    \underline{f}_v(k,l)=0, \underline{\mathcal{F}}_v(k,l)=\emptyset,  \underline{\mathcal{H}}_v(k,l)=\emptyset.\label{eq:zero_seeds_lower_f_outtree}
\end{equation}

Let $\mathcal{V}'_e \subseteq \mathcal{V}'$ be the set of leaf nodes\footnote{A leaf node is a node that does not have an outgoing neighboring node.} and $\mathcal{V}'_d \subseteq \mathcal{V}'$ be the set of dummy nodes in $G'$. Note that $\mathcal{V}'_e \cap \mathcal{V}'_d =\emptyset $ from the construction of $G'$. Then, for each $v\in \mathcal{V}'_e $, if $k\ge 1, l\ge 1$,
\begin{equation}
    \underline{f}_v(k,l)=0, \underline{\mathcal{F}}_v(k,l)=\emptyset,\underline{\mathcal{H}}_v(k,l)=\emptyset, \label{eq:leaf_physicalnode_notopen_outtree}
\end{equation}
\begin{multline}
    \overline{f}^a_v(k,l)=\overline{f}^b_v(k,l)=w_v,\overline{\mathcal{F}}^a_v(k,l)=\emptyset,\overline{\mathcal{F}}^b_v(k,l)=v,\\\overline{\mathcal{H}}_v^a(k,l)=\overline{\mathcal{H}}_v^b(k,l)=\overline{v}, \label{eq:leaf_sufficient_seeds_open_physical_nodes_outtree}
\end{multline}
where $\overline{v}$ is the physical node corresponding to $v$. Note that for each $v\in \mathcal{V}'_e $, if $k=0, l\ge 1$,
\begin{equation}
    \overline{f}^a_v(k,l)=w_v,\overline{\mathcal{F}}^a_v(k,l)=\emptyset,\overline{\mathcal{H}}_v^a(k,l)=\overline{v}. \label{eq:leaf_f^a_k=0}
\end{equation}

For each $v\in \mathcal{V}'_d$, when $k\ge 0, l=0$,
\begin{equation}
    \overline{f}^a_v(k,l)=0, \overline{\mathcal{F}}^a_v(k,l)=\emptyset,\overline{\mathcal{H}}^a_v(k,l)=\emptyset. \label{eq:zero_physical_nodes_upper_f_outtree}
\end{equation}

Let $\mathcal{L}_v=\{u_1,u_2\}$ be the set of outgoing neighbors of $v\in \mathcal{V}'$ and $\mathcal{V}'_i= \mathcal{V}'\setminus \mathcal{V}'_e$ be the set of internal nodes in $G'$. For each $v\in \mathcal{V}'_i$, $j\in\{1,2\} $, $0\le k_j\le K_s$ and $0\le l_j\le K_p$, let $g_{u_j}(k_j,l_j)=\max\{\underline{f}_{u_j}(k_j,l_j),\overline{f}_{u_j}^b(k_j,l_j)\}$ if $\overline{f}_{u_j}^b(k_j,l_j)$ is defined, otherwise $g_{u_j}(k_j,l_j)=\underline{f}_{u_j}(k_j,l_j)$. Then, for each $v\in \mathcal{V}'_i$, if $k\ge 1$, $l\ge 1$,
\begin{equation}
    \underline{f}_v(k,l) = \max_{k_1+k_2\le k; l_1+l_2\le l} \textstyle\sum_{j=1}^2  g_{u_j}(k_j,l_j), \label{eq:internalnode_twochildren_v_opened_outtree}
\end{equation}
\begin{equation}
    \{k_1^*,l_1^*,k_2^*,l_2^*\}\in \\ \argmax_{k_1+k_2\le k; l_1+l_2\le l} \textstyle\sum_{j=1}^2  g_{u_j}(k_j,l_j), \label{eq:internal_f_v_k_1^*_outtree}
\end{equation}
\begin{equation}
    \underline{\mathcal{F}}_v(k,l)=\cup_{j=1}^2 \mathcal{F}'_{u_j}(k_j^*,l_j^*),  \underline{\mathcal{H}}_v(k,l)=\cup_{j=1}^2 \mathcal{H}'_{u_j}(k_j^*,l_j^*), \label{eq:f_underscore_internal}
\end{equation}
where for  all $j\in \{1,2\}$, $\mathcal{F}'_{u_j}(k_j^*,l_j^*) =\overline{\mathcal{F}}^b_{u_j}(k_j^*,l_j^*)$, and  $\mathcal{H}'_{u_j}(k_j^*,l_j^*) =\overline{\mathcal{H}}^b_{u_j}(k_j^*,l_j^*)$ if
$\overline{f}_{u_j}^b(k_j^*,l_j^*)$ is defined and $\overline{f}_{u_j}^b(k_j^*,l_j^*) \ge \underline{f}_{u_j}(k_j^*,l_j^*)$, otherwise
$\mathcal{F}'_{u_j}(k_j^*,l_j^*) = \underline{\mathcal{F}}_{u_j}(k_j^*,l_j^*)$ and $\mathcal{H}'_{u_j}(k_j^*,l_j^*) = \underline{\mathcal{H}}_{u_j}(k_j^*,l_j^*)$.

Let $\mathcal{V}'_n=\mathcal{V}'\setminus \mathcal{V}'_d$ be the set of non-dummy nodes in $G'$. For each $v\in \mathcal{V}'_n\cap \mathcal{V}'_i$, when $k\ge 0$, $l\ge 1$,
\begin{equation}
    \overline{f}^a_v(k,l) = w_v+\max_{k_1+k_2\le k,l_1+l_2\le l-1} \textstyle \sum_{j=1}^2 f_{u_j}(k_j,l_j),  \label{eq:internalnode_twochildren_v_active_nondummy_outtree}
    \end{equation}
    \begin{equation}
    \{k_1^*,l_1^*,k_2^*,l_2^*\}\in\argmax_{k_1+k_2\le k,l_1+l_2\le l-1} \textstyle \sum_{j=1}^2 f_{u_j}(k_j,l_j), \label{eq:internal_nondummy_outtree_k_1^*_k^2*}
\end{equation}
\begin{equation}
\overline{\mathcal{F}}_v^a(k,l)= \cup_{j=1}^2\mathcal{F}_{u_j}(k_j^*,l_j^*),
    \overline{\mathcal{H}}_v^a(k,l)=\{\overline{v}\}\cup_{j=1}^2\mathcal{H}_{u_j}(k_j^*,l_j^*).  \label{eq:internal_nondummy_outtree_H_F}
\end{equation}
For each $v\in \mathcal{V}'_d\cap \mathcal{V}'_i$, when $k\ge 0$, $l\ge 1$, $ \overline{f}^a_v(k,l)$ is computed in the same way as when $v\in \mathcal{V}'_n\cap \mathcal{V}'_i$ except we set $l=l+1$ in the RHS of \eqref{eq:internalnode_twochildren_v_active_nondummy_outtree}-\eqref{eq:internal_nondummy_outtree_k_1^*_k^2*} and ensure $\overline{\mathcal{H}}_v^a(k,l)=\cup_{j=1}^2\mathcal{H}_{u_j}(k_j^*,l_j^*)$.

Also, for each $v\in \mathcal{V}'_n\cap \mathcal{V}'_i$, when $k\ge 1$, $l\ge 1$,
\begin{equation}
    \overline{f}^b_v(k,l) = w_v+\max_{k_1+k_2\le k-1,l_1+l_2\le l-1} \textstyle \sum_{j=1}^2 f_{u_j}(k_j,l_j),
\end{equation}
\begin{equation}
    \{k_1^*,l_1^*,k_2^*,l_2^*\}\in\argmax_{k_1+k_2\le k-1,l_1+l_2\le l-1} \textstyle \sum_{j=1}^2 f_{u_j}(k_j,l_j), 
\end{equation}
\begin{align}
\overline{\mathcal{F}}_v^b(k,l)&= \{v\}\cup_{j=1}^2 \mathcal{F}_{u_j}(k_j^*,l_j^*), \label{eq:internal_nondummy_outtree_just_F^b}\\
    \overline{\mathcal{H}}_v^b(k,l)&=\{\overline{v}\}\cup_{j=1}^2 \mathcal{H}_{u_j}(k_j^*,l_j^*). \label{eq:internal_nondummy_outtree_H^b_F^b}
\end{align}

For each $v\in \mathcal{V}'_i$ when $|\mathcal{L}_v|=1$, the above analysis holds by setting $k_2=0$ and $l_2 = 0$ in \eqref{eq:internalnode_twochildren_v_opened_outtree}-\eqref{eq:internal_nondummy_outtree_H^b_F^b}. 

\begin{algorithm} \caption{Dynamic Programming Algorithm} \label{alg:DP_out-tree}
Consider an out-tree $G'=\{\mathcal{V}',\mathcal{E}'\}$, where each node has at most two outgoing neighbors.
  \begin{algorithmic}[1] 
    \State For each node $v \in \mathcal{V}'_e$, initialize the values of $f_v(k,l)$, $ \mathcal{F}_v(k,l)$, $ \mathcal{H}_v(k,l), \forall k\ge 0,l\ge 0$ by \eqref{eq:f_v(k,l)_non_dummy_kge1_lge1_outtree}-\eqref{eq:leaf_f^a_k=0}.
    \State Repeat the following until the termination criterion is reached:
    \begin{itemize}
        \item Stop if there is no node $v\in \mathcal{V}'$ for which $f_{v}(k,l)$ has not been computed for all $k\ge 0$ and $l\ge 0$. Otherwise, arbitrarily select a node $v$ such that for all $k'\ge 0$ and $l'\ge 0$, $f_{v'}(k',l')$ has been computed for each outgoing neighbor $v'$ of $v$. Then, compute $f_{v}(k,l)$, $ \mathcal{F}_v(k,l)$, $ \mathcal{H}_v(k,l)$ for all $k$ and $l$ using \eqref{eq:f_v(k,l)_non_dummy_kge1_lge1_outtree}-\eqref{eq:zero_seeds_lower_f_outtree}, \eqref{eq:zero_physical_nodes_upper_f_outtree}-\eqref{eq:internal_nondummy_outtree_H^b_F^b}. 
    \end{itemize}
    \State Let $r$ be the root node of $\mathcal{V}'$. Then, set the seed nodes as the nodes in the set $ \mathcal{F}_{r}(K_s,K_p)$ and open the physical nodes in the set  $ \mathcal{H}_{r}(K_s,K_p)$. 
  \end{algorithmic} 
\end{algorithm}

Note that Algorithms \ref{alg:outtree_dummy_addition} and \ref{alg:DP_out-tree} are inspired from the paper \cite{wang2016bharathi}; however, the paper \cite{wang2016bharathi} focuses on influence maximization problem under \textit{DLTM} when the social network is a directed rooted tree with all the edges pointed towards the root node and thus does not consider the presence of physical nodes and heterogeneous weights for social nodes. Therefore, the presence of physical nodes and weighted social nodes make our problem more challenging and interesting as mentioned before.

We now present the main result of this section.
\begin{theorem} \label{thm:in-tree}
Suppose Assumptions \ref{assum:atleast_one_active_neighbor}, \ref{assum:no_secondary_edges} and \ref{assum:forest_out-tree} hold, i.e., there is a forest $G=\{\mathcal{V},\mathcal{E}\}$ of out-trees  with $M (\geq 1)$ social nodes, along with a set $\mathcal{W}$ of $M$ physical nodes. Then, Algorithm \ref{alg:DP_forest_outtree} is optimal.
\end{theorem}
\begin{IEEEproof}
Since the Steps 1 and 2 of Algorithm \ref{alg:DP_forest_outtree} generate an out-tree $G'$ from $G$ by adding dummy nodes, we first show that Algorithm \ref{alg:DP_out-tree} is optimal for graph $G'$ by arguing that \eqref{eq:f_v(k,l)_non_dummy_kge1_lge1_outtree}-\eqref{eq:internal_nondummy_outtree_H^b_F^b} hold. Note that \eqref{eq:f_v(k,l)_non_dummy_kge1_lge1_outtree}-\eqref{eq:f_overline_othercases} trivially hold, so we focus on the other conditions. In condition \eqref{eq:zero_seeds_lower_f_outtree}, $\underline{f}_v(k,l)=0$ because node $v$ is inactive from the definition of $\underline{f}_v(k,l)$ and thus it is not possible to activate other nodes in the out-tree of $v$ since either $k=0$ (i.e., there is no seed node in the out-tree of $v$) or $l=0$ (i.e., no physical node is opened in the out-tree of $v$). Also, $\underline{\mathcal{F}}_v(k,l)=\emptyset$ and $\underline{\mathcal{H}}_v(k,l)=\emptyset$ ensure that $ \underline{f}_v(k,l)=0 $, and thus \eqref{eq:zero_seeds_lower_f_outtree} holds.  
Note that \eqref{eq:leaf_physicalnode_notopen_outtree} (resp. \eqref{eq:leaf_sufficient_seeds_open_physical_nodes_outtree}) holds trivially since $v$ is an inactive (resp. active) leaf node; also note that while computing $\overline{f}^a_v(k,l)$ node $v$ is not set as a seed node but is set as a seed node while computing $\overline{f}^b_v(k,l)$ because of the definitions of $\overline{f}^a_v(k,l)$ and $\overline{f}^b_v(k,l)$. The condition \eqref{eq:leaf_f^a_k=0} follows in the same way as \eqref{eq:leaf_sufficient_seeds_open_physical_nodes_outtree}.

We now focus on the conditions for non-leaf (i.e., interior) nodes.
The condition \eqref{eq:zero_physical_nodes_upper_f_outtree} follows from the fact that no non-dummy node can be activated in the out-tree rooted at $v$ since $l=0$.
Note that since node $v$ is not active in $\underline{f}_v(k,l)$, functions $\overline{f}_{u_1}^a(k_1,l_1)$ and $\overline{f}_{u_2}^a(k_2,l_2)$ that assume node $v$ (which is the predecessor of $u_1$ and $u_2$) is active are not considered in the definitions of $g_{u_1}(k_1,l_1)$ and $g_{u_2}(k_2,l_2)$ in \eqref{eq:internalnode_twochildren_v_opened_outtree}-\eqref{eq:internal_f_v_k_1^*_outtree}
and thus \eqref{eq:internalnode_twochildren_v_opened_outtree}-\eqref{eq:f_underscore_internal} hold. The remaining conditions (i.e., \eqref{eq:internalnode_twochildren_v_active_nondummy_outtree}-\eqref{eq:internal_nondummy_outtree_H^b_F^b}) require that $v$ is an active non-dummy node (thus physical node $\overline{v}$ is opened and then at most $l-1$ physical nodes can be opened corresponding to the remaining social nodes in the out-tree rooted at $v$). Note that node $v$ is not selected as a seed node in \eqref{eq:internal_nondummy_outtree_H_F} but is selected as a seed node in \eqref{eq:internal_nondummy_outtree_just_F^b} due to the definitions of $\overline{f}^a_v(k,l)$ and $\overline{f}^b_v(k,l)$.  
Next, the computation of $\overline{f}^a_v(k,l)$ when $v\in \mathcal{V}'_d \cap \mathcal{V}'_i$, $k\ge 0$ and $l\ge 1$,  is similar to that when $v\in \mathcal{V}'_n \cap \mathcal{V}'_i$, $k\ge 0$ and $l\ge 1$ with the difference that there is no physical node corresponding to $v\in \mathcal{V}'_d$ and we thus set $l=l+1$ in the RHS of \eqref{eq:internalnode_twochildren_v_active_nondummy_outtree}-\eqref{eq:internal_nondummy_outtree_k_1^*_k^2*} and ensure $\overline{\mathcal{H}}_v^a(k,l)=\cup_{j=1}^2\mathcal{H}_{u_j}(k_j^*,l_j^*)$. Note that when node $v$ has only one outgoing neighbor $u_1$, then the conditions for the case when there are two outgoing neighbors would hold by not allocating any seed nodes and opening any physical nodes corresponding to the social nodes in the out-tree of $u_2$. Thus, \eqref{eq:f_v(k,l)_non_dummy_kge1_lge1_outtree}-\eqref{eq:internal_nondummy_outtree_H^b_F^b} hold and therefore Algorithm \ref{alg:DP_out-tree} computes an optimal solution for $G'$ from the definitions of $ \mathcal{F}_{r}(K_s,K_p)$ and $ \mathcal{H}_{r}(K_s,K_p)$. Note that an optimal solution for $G'$ is also optimal for $G$ because the weight of each dummy node is zero, a dummy node is not set as a seed node in $G'$ and there are no physical nodes corresponding to the dummy nodes. Thus, the result follows. 
\end{IEEEproof}
\begin{remark} \label{rem:complexity_forest_out-tree}
Note that Algorithm \ref{alg:DP_forest_outtree} is a polynomial-time algorithm because of the following. First, Algorithm \ref{alg:outtree_dummy_addition} has polynomial-time complexity because the outer loop in Step 1 of Algorithm \ref{alg:outtree_dummy_addition} is executed at most $O(N)$ times as there are at most $N$ nodes and the inner loop in the Step 1 of Algorithm \ref{alg:outtree_dummy_addition} is also executed at most $O(N)$ times as the maximum number of outgoing neighbors of a node is $N-1$. Thus, Step 1 of Algorithm \ref{alg:DP_forest_outtree} has polynomial-time complexity as there are at most $N$ out-trees in $G$. Step 2 of Algorithm \ref{alg:DP_forest_outtree} also has polynomial-time complexity as the loop in that step is executed $O(N)$ times (as there are at most $N$ root nodes). Finally, Step 3 of Algorithm \ref{alg:DP_forest_outtree} has polynomial-time complexity because of the following. Let $N'$ be the number of nodes in $G'$; note that $N'=O(N)$ from Steps 1 and 2 of Algorithm \ref{alg:DP_forest_outtree}. Then, the combined complexity of Steps 1 and 2 in Algorithm \ref{alg:DP_out-tree} is $O(N'K_s^2K_p^2)$ because there are $O(N'K_sK_p)$ parameters that need to be computed since $k\in \{0,K_s\},l\in \{0,K_p\}$, and in each computation of parameters such as $\underline{f}_v(k,l)$ and $\overline{f}_v(k,l)$, there are at most $O(K_sK_p)$ comparisons that need to be made;\footnote{Note that $O(K_sK_p)$ comparisons are required as each node $v\in \mathcal{V}'$ has at most two outgoing neighbors 
and that is why dummy nodes are added in Algorithm \ref{alg:DP_forest_outtree} to ensure that each node $v\in \mathcal{V}'$ has at most two outgoing neighbors.} note that $K_s=O(N)$ and $K_p=O(N)$ because $K_s\le N$ and $K_p\le N$.
Finally, Step 3 of Algorithm \ref{alg:DP_out-tree} takes $O(N')$ operations. 
\end{remark}

\begin{figure}[ht]
	\begin{center}
		\includegraphics[scale=0.4]{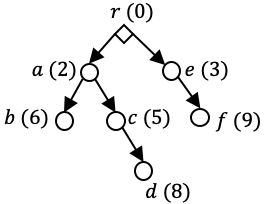}
	\end{center}	
	\caption{Graph obtained by adding a dummy node to the graph of Figure \ref{fig:non-trivial-forest_out-tree}.}
	\label{fig:modified_non-trivial-forest_out-tree}
\end{figure}

We now revisit Example \ref{exmp:forest_out-tree_non-trivial} to illustrate Algorithm \ref{alg:DP_forest_outtree}.
\begin{example}
Consider the instance of Problem \ref{problem} focused in Example \ref{exmp:forest_out-tree_non-trivial}. When Algorithm \ref{alg:outtree_dummy_addition} is applied to that example, graph $\overline{G}$ would be the same as graph $G$ in the first step of the algorithm as each node in $G$ has at most two outgoing neighbors. In the second step, a dummy node $r$ would be added as the root node as shown in Figure \ref{fig:modified_non-trivial-forest_out-tree} to form an out-tree $G'$. Finally, Algorithm \ref{alg:DP_out-tree} is run on $G'$ to obtain the optimal solution (where nodes $c$ and $f$ are selected as the seed nodes and the physical nodes corresponding to the social nodes $c,d$ and $f$ are opened).
\end{example}

\section{Conclusions}
In this paper, we studied the influence maximization problem in social networks that are dependent on a set of physical nodes. Each physical node covers one or more social nodes, and a necessary condition to activate a social node is that it should be covered by at least one of the opened physical nodes. There is a constraint on the total number of physical nodes that can be opened (in addition to the number of social nodes that can be chosen as seed nodes). This problem has applications in contexts such as disaster recovery where a displaced social group may decide to return to its home only if some infrastructure components in its residential neighborhood have been repaired and a sufficiently large number of groups in its social network have returned back. The general problem is NP-hard to approximate within any constant factor and therefore we provided optimal and approximation algorithms for special instances of the problem. 

There are several avenues of potential future research along the lines of our work. 
In this paper, we characterized algorithms for cases of Problem \ref{problem} under the Assumptions \ref{assum:atleast_one_active_neighbor} and \ref{assum:no_secondary_edges}, and therefore characterizing approximation algorithms under more general conditions would be of interest. Considering dependencies between the physical nodes would also be an important extension. 
\section{Acknowledgement}
We thank Dr. Kent Quanrud and Dr. Seungyoon Lee for their guidance on this work. This research was supported by National Science Foundation (NSF) grant CMMI 1638311. 

\bibliographystyle{IEEEtran}
\bibliography{refs}

\end{document}